\documentclass[12pt]{article} %
\usepackage[sectionbib]{natbib}
\usepackage{array, epsfig, fancyheadings, rotating}
\usepackage{float} 
\usepackage[]{hyperref}  
\usepackage{sectsty,  secdot}
\sectionfont{\fontsize{12}{14pt plus.8pt minus .6pt}\selectfont}
\renewcommand{\theequation}{\thesection\arabic{equation}}
\subsectionfont{\fontsize{12}{14pt plus.8pt minus .6pt}\selectfont}

\usepackage{booktabs}
\usepackage{tikz}
\usetikzlibrary{arrows.meta}

\usepackage{float}
\usepackage{graphicx}
\usepackage{subcaption}
\usepackage{lscape}
\usepackage{pdflscape} 
\usepackage{longtable}
\usepackage{threeparttable}
\usepackage{threeparttablex}
\usepackage{enumitem}
\usepackage{amsmath}
\usepackage{amssymb}
\usepackage{amsfonts}
\usepackage{multirow}
\usepackage{amsthm}
\usepackage{indentfirst}
\usepackage{graphicx}
\usepackage{tikz}
\usetikzlibrary{arrows}          
\usetikzlibrary{arrows.meta}     
\usetikzlibrary{positioning}     
\usepackage{tikz}
\usepackage{pgfplots}
\pgfplotsset{compat=1.18} 
\newtheorem{corollary}{Corollary}
\newtheorem{proposition}{Proposition}
\theoremstyle{definition}

\newtheoremstyle{myremark}
{12pt}
{12pt}
{\itshape}
{}
{\bfseries}
{.}
{5mm}
{\thmname{#1}\thmnumber{ #2}\thmnote{ (#3)}} 
\theoremstyle{myremark}

\usepackage{algorithm}
\usepackage{algorithmic}
\DeclareMathOperator{\sgn}{sgn}

\usepackage{xr}
\makeatletter
\newcommand*{\addFileDependency}[1]{
  \typeout{(#1)}
  \@addtofilelist{#1}
  \IfFileExists{#1}{}{\typeout{No file #1.}}
}
\makeatother
\def\var{\operatorname{Var}}

 \newtheorem{innercustomgeneric}{\customgenericname}
\providecommand{\customgenericname}{}
\newcommand{\newcustomtheorem}[2]{%
  \newenvironment{#1}[1]
  {%
   \renewcommand\customgenericname{#2}%
   \renewcommand\theinnercustomgeneric{##1}%
   \innercustomgeneric
  }
  {\endinnercustomgeneric}
}
\newcustomtheorem{customthm}{Proposition}
\newcustomtheorem{customlemma}{Lemma}
\newcustomtheorem{customcorollary}{Corollary}

 \usepackage{algorithm}
\usepackage{algorithmic}

\begin{document}

\renewcommand{\baselinestretch}{2}



\markright{ \hbox{\footnotesize\rm Statistica Sinica
}\hfill\\[-13pt]
\hbox{\footnotesize\rm
}\hfill }

\markboth{\hfill{\footnotesize\rm FIRSTNAME1 LASTNAME1 AND FIRSTNAME2 LASTNAME2} \hfill}
{\hfill {\footnotesize\rm FILL IN A SHORT RUNNING TITLE} \hfill}

\renewcommand{\thefootnote}{}
$\ $\par


\fontsize{12}{14pt plus.8pt minus .6pt}\selectfont \vspace{0.8pc}
\vspace{.4cm}\centerline{\large\bf Bidirectional causal inference for binary outcomes  }
\vspace{.2cm} \centerline{\large\bf in the presence of unmeasured confounding}\vspace{.4cm} 
 \centerline{Yafang Deng$^{1}$ ,  Kang Shuai$^{2}$,  Shanshan Luo$^{1}$  }
\vspace{.4cm} 
\centerline{\it \textsuperscript{1} School of Mathematics and Statistics, }
\centerline{\it   Beijing Technology and Business University}
		\centerline{\it 	\textsuperscript{2} The Dartmouth Institute for Health Policy and Clinical Practice, }
        \centerline{\it  Geisel School of Medicine at Dartmouth } 
 \vspace{.55cm} \fontsize{9}{11.5pt plus.8pt minus.6pt}\selectfont


  
\begin{quotation}
 \noindent {\it Abstract:}
Bidirectional causal relationships arising from mutual interactions between variables are commonly observed within biomedical, econometrical, and social science contexts. When such relationships are further complicated by unobserved factors,  identifying causal effects in both directions becomes especially challenging.
 For continuous variables, methods that utilize two instrumental variables from both directions have been proposed to explore bidirectional causal effects within linear models. However, the existing techniques are not immediately applicable when the key variables of interest are binary. 
{To address these issues},  we propose a structural equation modeling approach that links observed binary variables to latent continuous variables through a constrained mapping.   { We firstly establish the identification results for bidirectional causal effects using a pair of instrumental variables. Then, we develop an estimation method for the corresponding causal parameters. We also conduct sensitivity analysis under scenarios where certain identification conditions are violated.} Finally, we apply our approach to investigate the bidirectional causal relationship between heart disease and diabetes, demonstrating its practical utility in biomedical researches.

 \vspace{9pt}

 \par
 \end{quotation}\par

\makeatletter
\newcommand{\indep}{%
 \mathbin{%
  \mathpalette{\@indep}{}%
 }%
}
\newcommand{\nindep}{%
 \mathbin{
  \mathpalette{\@indep}{\not}
 }%
}
\newcommand{\@indep}[2]{%
 \sbox0{$#1\perp\m@th$}
 \sbox2{$#1=$}
 \sbox4{$#1\vcenter{}$}
 \rlap{\copy0}
 \dimen@=\dimexpr\ht2-\ht4-.2pt\relax
 \kern\dimen@
 {#2}%
 \kern\dimen@
 \copy0 
}

\newtheorem{model}{Model}
\newtheorem{assumption}{Assumption}
\newtheorem{condition}{Condition}  


\def\var{\mathrm{var}}
\def\mix{\mathrm{mix}} 
\def\AT{{\mathrm{AT}}}
\def\CO{{\mathrm{CO}}}
\def\Pr{{\mathrm{pr}}}
\def\wls{{\mathrm{hjk}}}
\def\ipw{{\mathrm{ipw}}}
\def\EIF{{\mathrm{EIF}}}
\def\NT{{\mathrm{NT}}}
\def\DE{{\mathrm{DE}}}
\def\ATE{{\mathrm{ATE}}}
\def\ai{\mathscr{AI}}
\def\bone{\mbox{\bo10math $1$}}
\def\bzero{\mbox{\bo10math $0$}}
\def\bo10z{\mbox{\bo10math $Z$}}
\def\lbo10z{\mbox{\bo10math $z$}}
\def\bo10y{\mbox{\bo10math $Y$}}
\def\lbo10y{\mbox{\bo10math $y$}}
\def\bo10u{\mbox{\bo10math $u$}}
\def\E{\mathbb{E}}
\def\Ber{\mathrm{Ber}}
\def\Bern{\mathrm{Bern}}
\def\full{\mathrm{full}}
\def\obs{\mathrm{obs}}
\def\pr{\textnormal{pr}}
\def\cau{\textnormal{c}}
\def\s{\textnormal{s}}
\def\Pro{\textnormal{P}}
\def\P{\mathbb{P}}
\def\bo10d{\mbox{\bo10math $d$}}

\newcommand{\red}{\color{red}}
\newcommand{\prered}{\color{red}}
\newcommand{\blue}{\color{blue}}
\newcommand{\MR}{\textnormal{tr}}
\newcommand{\DR}{\textnormal{dr}}
\def\O{\mathcal{O}}
\allowdisplaybreaks
\newcommand\Pn{{\mathbb{P}_n}}
\def\T{{ \mathrm{\scriptscriptstyle T} }}

\def\thefigure{\arabic{figure}}
\def\thetable{\arabic{table}}

\renewcommand{\theequation}{\thesection.\arabic{equation}}

\fontsize{12}{14pt plus.8pt minus .6pt}\selectfont

\section{Introduction}\label{section:introduction}
The emergence of feedback mechanisms between entities often signifies mutual causal influences, a ubiquitous phenomenon across real-world systems \citep{Smith1999,Engelgau2019,Riaz2017}.
 In economics, a bidirectional causal relationship exists between health status and wealth accumulation: poor health erodes economic resources, while economic deprivation compromises health, forming a self-reinforcing cycle \citep{Smith1999,Engelgau2019}. In biomedicine, tumor cells and the immune system engage in mutual causation—immune surveillance exerts selective pressures that sculpt cancer clonal evolution, whereas cancer cells deploy immunosuppressive molecules to subvert immune responses, establishing a malignant ``immunoediting-escape'' loop \citep{Riaz2017}. Similarly, the gut microbiota and brain maintain bidirectional communication via neuro-immune-endocrine pathways: stress-induced neural activity alters microbial composition, and dysbiosis independently disrupts brain function \citep{Foster2017,Cryan2019}. Psychologically, sleep and depression exhibit reciprocal causality: chronic insomnia elevates risks for psychiatric disorders, while over 90\% of depressed patients manifest sleep disruptions consistent with insomnia \citep{Palagini2013}. All these practical phenomenon highlights the significance of investigating reverse causality, and methods for bidirectional causal effects definitely require further exploration.

However, in the study of bidirectional causal relationships, the presence of unobserved confoundings complicates the real-world problems. It may come with confounding bias so that the true causal effect cannot be recovered, which leads to invalid causal conclusions.  
To address this issue, instrumental variable (IV) approach {offers a potential solution for identifying causal effects by eliminating the confounding bias.} The related IV researches on bidirectional causal inference are primarily conducted within Mendelian Randomization (MR) framework. {Specifically,} MR utilizes independent genetic IVs for two traits to effectively {disentangle the causal relations} between them, thereby {separating the additional} associations caused by reverse causation \citep{DaveySmith2014}. For example,  \citet{gao2019bidirectional} regarded reverse causality as a source of confounding and used two-sample MR to explore the causal relationships between insomnia and five major psychiatric disorders. \citet{Xue2020} refined the traditional bidirectional MR approach, enabling robust inference of causal direction between two traits even in the presence of horizontal pleiotropy. \citet{li2024focusing} assessed commonly used identifiability assumptions in bidirectional MR models and proposed a new focusing framework to test bidirectional causal effects with potentially invalid instruments. \citet{Xie2024BiDirectMR} addressed the identifiability within bidirectional MR models by correctly selecting valid instrument sets using observational data and developed a cluster fusion-like method to discover valid instruments.

Previous research on bidirectional causal inference based on MR primarily focuses on recovering causal effects and related applications with continuous outcomes. However, binary outcomes are also frequently encountered in applied settings, particularly in epidemiological studies involving the disease risk factors. For instance,  \citet{tenhave2003causal} conducted a placebo-controlled randomized trial on cholesterol levels, where participants in the treatment group listened to audio tapes delivering dietary and nursing interventions, and the binary outcome was defined as whether or not participants had high cholesterol levels. Similarly,  \citet{Borracci2013} evaluated the impact of preoperative oral morphine sulfate on postoperative pain relief,  where two binary outcome variables were constructed based on resting and movement pain scores using established thresholds  \citep{Lupparelli2020}.  
Compared with continuous variables,  simple linear models are not applicable for binary outcomes. 
\citet{clarke2012instrumental} provided a comprehensive review of IV estimation methods for binary outcomes, focusing on identification strategies that rely on linear selection models and the control function approach. However, these methods have thus far been applied exclusively within traditional unidirectional causal frameworks.


In this paper, we propose bidirectional causal frameworks under unmeasured confounding for observed binary variables, employing IV approaches. 

This paper makes the following three main contributions: {First, we formulate the causal problem based on a linear structural equation modeling (SEM) framework to capture bidirectional causal relationships between continuous latent variables.}   {  Second, we use a pair of IVs to establish the identifiability of bidirectional causal effects and derive their estimators, along with the corresponding asymptotic properties.} {Third, we perform sensitivity analysis to evaluate the proposed method when key assumptions are possibly violated, including the exclusion restriction assumption and the restrictive normality assumption.}

{The remainder of this paper is organized as follows. Section~\ref{section:methodology} presents the proposed causal framework, including the problem formulation, core assumptions and the identification results. Section~\ref{section:sensitivity} conducts sensitivity analysis for two main assumptions, including the normality and exclusion restriction assumptions. To illustrate the numerical performance of our method, in Section~\ref{section:simulation}, we conduct detailed simulation experiments under various settings when the key conditions are met or not met. We further apply the proposed method to study the relationship between heart disease and diabetes in Section~\ref{section:case}. Finally, we briefly concludes this paper in Section~\ref{section:summary}.}

\section{Notation and Framework}\label{section:methodology}
\subsection{Binary bidirectional model}\label{subsec:binary model}

Let $X \in \{0, 1\}$ be a binary treatment variable and $Y \in \{0, 1\}$ an observed binary outcome. In observational studies, both $X$ and $Y$ may be influenced by background covariates, including observed covariates $C$ and unobserved confounders $U$ and $V$. In our motivating application, $X$ indicates whether one has coronary heart disease or not, and $Y$ indicates whether one has diabetes or not. Previous studies~ \citep{Henning2018, Katakami2018} have shown that individuals with diabetes ($Y = 1$) are at a significantly high risk of having heart disease ($X = 1$); conversely, complications or treatments associated with heart disease ($X = 1$) may impair glucose regulation, thereby increasing the risk of  diabetes ($Y = 1$).

In such scenarios, assuming a unidirectional causal relationship (e.g., $X \rightarrow Y$) may ignore potential reverse effects, leading to biased estimates of causal effects. To more accurately capture the mutual dependence between the variables, we introduce a bidirectional causal mechanism to account for the possible reciprocal influences. Specifically, we assume the following generating mechanism  \citep{hausman1983specification}:
\begin{equation}
    \label{eq:selection-bidirection}
    \begin{aligned}
X^\circ &= \mu_{x0}+ \beta_{y \to x} Y^\circ +  \mu_{xc} C +   U, \quad X = \mathbb{I}(X^\circ > 0), \\
Y^\circ &= \mu_{y0} + \beta_{x \to y} X^\circ + \mu_{yc} C  + V,  \quad Y = \mathbb{I}(Y^\circ > 0),
\end{aligned}
\end{equation}
where $C\indep (U,V)$, $X^\circ$ and $Y^\circ$ represent the latent continuous variables corresponding to the observed binary outcomes $X$ and $Y$, and $\mathbb{I}(\cdot)$ is the indicator function that maps the latent variables to binary responses. { We have provided a discussion on the presence of the latent continuous variables in Section E of the Supplementary Materials.}

 {The structural equation \eqref{eq:selection-bidirection} characterizes the bidirectional causal relationship at equilibrium. We provide a detailed discussion of the bidirectional causal mechanism and the causal interpretation of the parameters $\beta_{x\to y}$ and $\beta_{y\to x}$ in Sections~F and~G of the Supplementary Materials.} When $\beta_{y \to x} = 0$, Model~\eqref{eq:selection-bidirection} reduces to a unidirectional causal model. 

Model \eqref{eq:selection-bidirection} presents substantial challenges due to the coexistence of the bidirectional causation, latent variable structures, and unobserved confounding. First, the reciprocal dependence between the latent variables $X^\circ$ and $Y^\circ$ forms a simultaneous equation system, complicating identification and estimation, as standard regression techniques fail to { recover true causal effects}. Second, the observed binary variables $X$ and $Y$ are thresholded versions of continuous latent variables, which necessitates additional distributional assumptions to characterize the dependence structure between the observed variables and latent variables. Third, the unobserved confounders $U$ and $V$ that simultaneously affect both equations introduce endogeneity, leading to biased estimates unless properly addressed.

To establish identification results, the next subsection introduces the structural and IV assumptions. These assumptions form the basis for the estimation procedures developed in this paper. Section \ref{section:sensitivity} further conducts sensitivity analyses to examine the robustness of our conclusions to possible violations of key assumptions.

\subsection{Assumptions}\label{subsec:assumptions}
\begin{figure}[htbp]
\centering
\captionsetup{font=small}
\captionsetup[subfigure]{aboveskip=2pt, belowskip=2pt, justification=centering}

\begin{subfigure}[t]{0.45\textwidth}
\centering
\resizebox{\linewidth}{!}{%
\begin{tikzpicture}[>=latex, scale=0.7]
\tikzstyle{node} = [draw, circle, thick, 
                            minimum size=1cm,
                            inner sep=1pt,
                            text width=0.8cm, 
                            align=center]

\node[node] (Z) at (-3.3,0) {$Z$};
\node[node] (X_) at (0,0) {$X^\circ$};
\node[node] (X) at (0,-2.8) {$X$};
\node[node] (Y_) at (3.5,0) {$Y^\circ$};
\node[node] (Y) at (3.5,-2.8) {$Y$};
\node[node, fill=gray!30] (U) at (0,2.75) {$U$}; 
\node[node, fill=gray!30] (V) at (3.5,2.75) {$V$};
\node[node] (W) at (6.8,0) {$W$};

\draw[thick, ->] (Z) -- (X_);
\draw[thick, ->] (X_) -- (X);
\draw[thick, ->] (Y_) -- (Y);
\draw[thick, ->] (X_) to[out=20,in=160] (Y_);
\draw[thick, ->] (Y_) to[out=200,in=340] (X_);
\draw[thick, ->] (U) -- (X_);
\draw[thick, - ] (U) -- (V);
\draw[thick, ->] (V) -- (Y_);
\draw[thick, ->] (W) -- (Y_);
\end{tikzpicture}}
\caption{  }
\label{fig:bidirectional_graph}
\end{subfigure}
\hfill
\begin{subfigure}[t]{0.45\textwidth}
\centering
\resizebox{\linewidth}{!}{%
\begin{tikzpicture}[>=latex, scale=0.7]
\tikzstyle{node} = [draw, circle, thick, 
                            minimum size=1cm,
                            inner sep=1pt,
                            text width=0.8cm, 
                            align=center]

\node[node] (Z) at (-3.5,0) {$Z$};
\node[node] (X_) at (0,0) {$X^\circ$};
\node[node] (X) at (0,-2.75) {$X$};
\node[node] (Y_) at (3.5,0) {$Y^\circ$};
\node[node] (Y) at (3.5,-2.75) {$Y$};
\node[node, fill=gray!30] (U) at (0,2.75) {$U$}; 
\node[node, fill=gray!30] (V) at (3.5,2.75) {$V$};
\node[node] (W) at (7,0) {$W$};

\draw[thick, ->] (Z) -- (X_);
\draw[thick, ->] (X_) -- (X);
\draw[thick, ->] (Y_) -- (Y);
\draw[thick, ->] (X_) to[out=20,in=160] (Y_);
\draw[thick, ->] (Y_) to[out=200,in=340] (X_);
\draw[thick, ->] (U) -- (X_);
\draw[thick, ->] (V) -- (Y_);
\draw[thick, ->] (W) -- (Y_);
\end{tikzpicture}}
\caption{ }
\label{fig:bidirectional_given_assu_1}
\end{subfigure}
\caption{Graphic illustration of Model~\eqref{eq:bidirectional_model} with IVs $Z$ and $W$.}
\label{fig:fig1}
\end{figure}
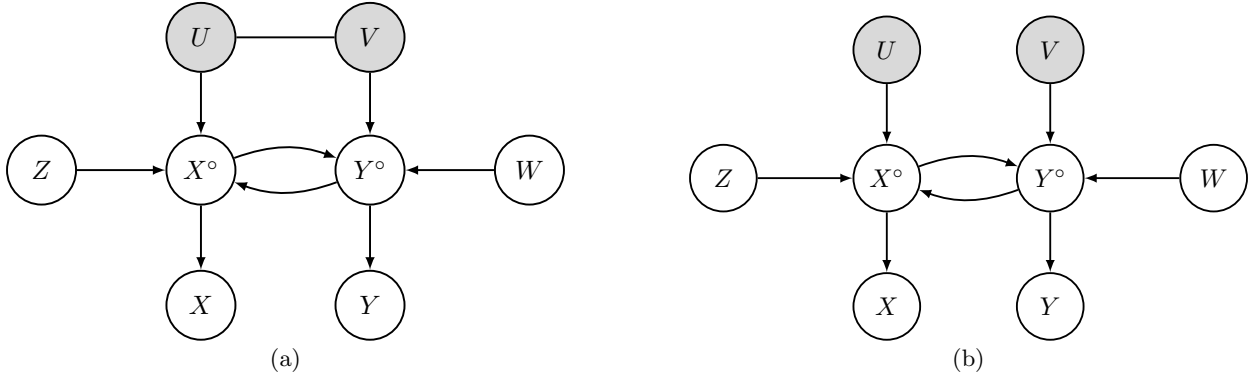

The covariate vector $C$ includes two IVs, $Z$ and $W$, each influencing a single causal direction. Specifically, $Z$ only directly affects $X^\circ$ and $W$ only directly affects $Y^\circ$. The remaining components of $C$, denoted by $Q$, represent other observed covariates that may impact both $X$ and $Y$. For notational simplicity, we omit $Q$ in the following discussion. Figure~\ref{fig:bidirectional_graph} illustrates the structure of the bidirectional causal framework for binary variables with IVs $Z$ and $W$.  By incorporating $Z$ and $W$, the bidirectional causal Model \eqref{eq:selection-bidirection} can be reformulated as the following form:
\begin{align}\label{eq:bidirectional_model}
    \begin{gathered}
X^\circ = \mu_{x0}+ \beta_{y \to x} Y^\circ + \mu_{xz} Z  +  U  , ~~ X =\mathbb{I}\left(X^\circ>0\right),\\
    Y^\circ = \mu_{y0} + \beta_{x \to y} X^\circ + \mu_{yw} W +  V , ~~ Y =\mathbb{I}\left(Y^\circ>0\right).
\end{gathered}
\end{align}
 The parameters $\mu_{xz}$ and $\mu_{yw}$ in Model~\eqref{eq:bidirectional_model} characterize the effects of $Z$ and $W$ on the latent variables $X^\circ$ and $Y^\circ$,  respectively. From Model~\eqref{eq:bidirectional_model}, it is clear that $Z$ and $W$ serve as valid IVs for each causal direction.  Throughout this paper, we maintain this terminology for IV in the bidirectional framework because they satisfy the exclusion restriction (ER) by having no direct effects on the outcome variables, analogous to their role in traditional unidirectional models  \citep{Angrist1996}. This terminology is consistent with other bidirectional causal inference literature {\citep{Xue2020,li2024focusing,Xie2024BiDirectMR}.}

In our data illustration, the binary variables $X$ and $Y$ in Figure~\ref{fig:bidirectional_graph} represent two observed binary indicators: $X$ indicates whether an individual has coronary heart disease (or myocardial infarction), and $Y$ indicates whether an individual has diabetes. The variables $X^\circ$ and $Y^\circ$ correspond to their underlying continuous latent variables. 
In this context,  “whether an individual has had a stroke” can be used as the IV $Z$ for $X^\circ$,  and “body mass index (BMI)” can be used as the IV $W$ for $Y^\circ$. This choice is based on the following knowledge: stroke will possibly have direct effect on heart disease but unlikely has a direct influence on diabetes, while BMI can directly affects diabetes and is less likely to directly impact the heart disease. 
Also, the covariates $Q$ can represent other background characteristics such as age, gender, and race, while $U$ and $V$ denotes unobserved confounding factors not captured in the dataset, such as genetic predisposition and lifestyle factors. Based on our setting, we need to further impose the following normality assumption to achieve identifiability.

    \begin{assumption}[Restrictive normality]\label{assu:U}
    (i)  $V,U\sim N(0,\sigma^2)$, { where $\sigma$ is an unknown constant,} (ii) $\mathrm{Cov}(V,U)=0$.
\end{assumption}

Assumption~\ref{assu:U} imposes the distributional restrictions on $U$ and $V$ to facilitate identification and simplify the modeling framework. Assumption~\ref{assu:U}(i) assumes that $U$ and $V$ both follow normal distributions with zero means and equal variances. {Assumption~\ref{assu:U}(ii) further assumes that $U$ and $V$ are uncorrelated, i.e., mutually independent in the normal setting, and the endogeneity in the bidirectional causal model~\eqref{eq:bidirectional_model} for each direction arises only from reverse causality.}  Figure~\ref{fig:bidirectional_given_assu_1} shows the causal diagram that satisfies Assumption~\ref{assu:U}. This assumption is primarily introduced to simplify computation. Under this framework, {Section~\ref{ssec:Identification} discusses the identification of causal effects within the bidirectional causal framework.  Additionally, Section~\ref{ssec:equal-confounding} presents the sensitivity analysis results under more general settings, { evaluating the possible violations of Assumption~\ref{assu:U}.}
}

{ Before formally introducing the identification results, we firstly rewrite the structural  equation  Model~\eqref{eq:bidirectional_model} in the following reduced form when $\beta_{x \to y} \beta_{y \to x} \neq 1$:}
\begin{equation}\label{eq:identify_start}
   \begin{aligned}
        X^\circ = \theta_{x0}+  \theta_{xz} Z+  \theta_{xw} W  +\theta_{xu} U+\theta_{xv} V  , ~~ X =\mathbb{I}\left(X^\circ>0\right),\\
    Y^\circ = \theta_{y0} +  \theta_{yz} Z+  \theta_{yw} W +\theta_{yu} U +\theta_{yv} V , ~~ Y =\mathbb{I}\left(Y^\circ>0\right),
   \end{aligned}
\end{equation}
where $c = 1/(1-\beta_{x\rightarrow y}\beta_{y\rightarrow x})$, $\theta_{x0} = c(\mu_{x0}+\beta_{y\rightarrow x}\mu_{y0})$, $\theta_{xz} = c\mu_{xz}$, $\theta_{xw} = c\beta_{y\rightarrow x}\mu_{yw}$, $\theta_{xu} = c$, $\theta_{xv} = c\beta_{y\rightarrow x}$; $\theta_{y0} = c(\beta_{x\rightarrow y}\mu_{x0}+\mu_{y0})$, $\theta_{yz} = c\beta_{x\rightarrow y}\mu_{xz}$, $\theta_{yw} = c\mu_{yw}$, $\theta_{yu}=c\beta_{x\rightarrow y}$, and $\theta_{yv} = c$.

\subsection{Identification}\label{ssec:Identification}
Under the structural assumptions and Assumption~\ref{assu:U}, the conditional probabilities of the two binary variables $X$ and $Y$ given the two observed IVs $Z$ and $W$ can be derived through their latent continuous representations $X^\circ$ and $Y^\circ$. Under Model~\eqref{eq:identify_start},   we can derive the two observed  conditional probabilities as follows, with $\Phi(\cdot)$ being the cumulative distribution function (CDF) of standard normal distribution:
\begin{align}\label{eq:probit_for_assu1}
\begin{aligned}
    \mathrm{pr}(X = 1 \mid Z, W ) 
= \mathrm{pr}(X^\circ > 0 \mid Z, W ) 
= \Phi\left( 
\xi_{x0} + \xi_{xz} Z + \xi_{xw} W 
\right),\\\mathrm{pr}(Y = 1 \mid Z, W ) 
= \mathrm{pr}(Y^\circ > 0 \mid Z, W ) 
= \Phi\left( 
\xi_{y0} + \xi_{yz} Z + \xi_{yw} W
\right),
\end{aligned}
\end{align} 
where $\lambda_1 = \mathrm{Var}(\theta_{xu}U+\theta_{xv}V)$,
$ 
\xi_{x0} =  {\theta_{x0}}/\sqrt{\lambda_1 }$, $  
\xi_{xz} =  {\theta_{xz}}/\sqrt{\lambda_1 }$, $  
\xi_{xw} =  {\theta_{xw}}/\sqrt{\lambda_1 }$, and  $\lambda_2$, $ \xi_{y0}$, $\xi_{yz}$, and $ \xi_{yw}$ are similarly denoted. Expressions in Model~\eqref{eq:probit_for_assu1} depend only on the observed variables $Z$ and $W$, allowing us to estimate the parameters  directly using observational data.  We summarize the identification results in Proposition~\ref {proposition:identify}.

\begin{proposition}\label{proposition:identify}
Under Model~\eqref{eq:bidirectional_model}  and Assumption~\ref{assu:U},  the bidirectional causal effects  are identifiable:
\begin{align} \label{eq:beta_identify}
\beta_{x \to y} = \frac{\xi_{yz}}{\xi_{xz}} \sqrt{\frac{\xi_{xw}^2 \xi_{xz}^2 - \xi_{yw}^2\xi_{xz}^2}{\xi_{yz}^2\xi_{yw}^2 - \xi_{yw}^2\xi_{xz}^2}},~~
\beta_{y \to x} = \frac{\xi_{xw}}{\xi_{yw}} \sqrt{\frac{\xi_{yz}^2 \xi_{yw}^2 - \xi_{xz}^2\xi_{yw}^2}{\xi_{xw}^2\xi_{xz}^2 - \xi_{xz}^2\xi_{yw}^2}}. 
\end{align}
where the terms $\xi_{xz}$,  $\xi_{yz}$,  $\xi_{xw}$,  and $\xi_{yw}$ are the   coefficients in Model~\eqref{eq:probit_for_assu1}.
\end{proposition}

The detailed proof of Proposition~\ref{proposition:identify} is provided in Section A.1 of the {Supplementary Materials}. Proposition~\ref{proposition:identify} gives the identification formulas for the causal parameters $\beta_{x \to y}$ and $\beta_{y \to x}$ based on the identifiable coefficients from Model~\eqref{eq:probit_for_assu1}. As shown in Equation~\eqref{eq:beta_identify}, the introduction of two IVs is essential, as they differentially affect the binary variables $X$ and $Y$. This difference is captured through the coefficients $\xi_{xz}$, $\xi_{yz}$, $\xi_{xw}$, and $\xi_{yw}$ in Probit models. 
 In addition, assuming zero covariance between $U$ and $V$ in Assumption~\ref{assu:U} simplifies the parameters in Model~\eqref{eq:probit_for_assu1}, yielding a symmetric and simple identification formula  \eqref{eq:beta_identify}.  
Proposition~\ref{proposition:identify} can be interpreted as the identification result when the two unobserved confounders are uncorrelated. { Furthermore, it can be seen from Equation~\eqref{eq:beta_identify} that the omission of the baseline covariate $Q$ does not affect identifiability. In the numerical simulations in Section~\ref{section:simulation}, we will conduct experiments based on the full model that includes the baseline covariate.} In Section~\ref{ssec:equal-confounding}, we will extend this framework to the scenario depicted in Figure~\ref{fig:bidirectional_graph}, which incorporates the correlation between confounders, thereby deriving more general identification results. Furthermore, we provide Algorithm S1 for estimation in Section A.2 of the Supplementary Materials and establish asymptotic properties in Proposition S1, with the corresponding proofs given in Section A.3.

{ We further consider extending Assumption~\ref{assu:U} to non-Gaussian cases and conducting a robustness analysis of Assumption~\ref{assu:U}. For details, see Section~H of the Supplementary Materials. }

\section{Sensitivity Analysis}\label{section:sensitivity}
\subsection{Sensitivity analysis for Assumption~\ref{assu:U}}\label{ssec:equal-confounding}
In Section~\ref{ssec:Identification}, we have introduced a strong normality Assumption~\ref{assu:U} for identification. Specifically, this assumption requires that (i) the unobserved confounders $U$ and $V$ follow normal distributions with mean zero and equal variance $\sigma^2$; and (ii) they are independent, i.e., $\mathrm{Cov}(U,V)=0$. These two requirements are rather restrictive, since in practice it is often unrealistic to assume that the unobserved confounders in the two models have identical variances and are completely independent. 
In this subsection, while still utilizing the IVs, we conduct a sensitivity analysis for Assumption~\ref{assu:U}.  We propose the following new assumption that is weaker than Assumption~\ref{assu:U}, allowing $U$ and $V$ to have different variances and be potentially correlated.

\begin{assumption}[Normality]\label{assu:U-2}
(i) $V \sim {N}(0,\sigma^2)$ and  $U \sim {N}(0,\gamma_1\sigma^2)$, 
(ii) $\mathrm{Cov}( V,U) =  \gamma_2\sigma ^2$. { The parameter $\sigma$ is an unknown constant, while $\gamma_1$ and $\gamma_2$ are sensitivity parameters.}

\end{assumption}

To quantify the influence of the relationship between unobserved confounders $U$ and $V$, we introduce two sensitivity parameters in Assumption~\ref{assu:U-2}:
{$
\gamma_1 = \mathrm{Var}(U)/ \mathrm{Var}(V)$   {and}  $  \gamma_2 = {\mathrm{Cov}(U,V)}/{\mathrm{Var}(V)}.
$}  
{ The parameter $\gamma_1$ measures the relative strength of the variances of $U$ and $V$, $\gamma_2$ characterizes the ratio of the covariance of $(U, V)$ to the variance of $V$, and the presence of $\gamma_2$ implies that the endogeneity in model~\eqref{eq:bidirectional_model} arises from both reverse causality and unobserved confounders. We will examine the robustness of the proposed method by varying these two sensitivity parameters.} It is clear that the two sensitivity parameters satisfy ${\gamma_1 } \geq \gamma_2 ^2$. When $\gamma_1 = 1$ and $\gamma_2 = 0$, Assumption~\ref{assu:U-2} is equivalent to Assumption~\ref{assu:U}. The two parameters enable us to investigate how the relative variances of $U$ and $V$, together with their covariance $\mathrm{Cov}(U,V)$, influence the identification and estimation of the bidirectional causal effects.

In this context, we introduce two notations: $k_1 \equiv \xi_{yz}/\xi_{xz}$ and $k_2 \equiv \xi_{xw}/\xi_{yw}$, to simplify subsequent discussion. They represent the ratios of Probit regression coefficients of $X$ and $Y$ on $(Z, W)$, respectively. The following proposition provides the identification results:
  { 
\begin{proposition}\label{prop:UV}
Under Model~\eqref{eq:bidirectional_model} and
Assumption~\ref{assu:U-2}, given the sensitivity parameters
$\gamma_1$ and $\gamma_2$, the bidirectional causal effects can be obtained through the following two candidate solution pairs:
\begin{equation}\label{eq:identify for UV}
\begin{aligned}
\beta_{x\rightarrow y,1}
&=
\frac{\gamma_2k_1(k_2-k_1)+|k_1|\sqrt{\Delta_1}}
{\gamma_1(k_1^2-1)},
&
\beta_{y\rightarrow x,1}
&=
\frac{\gamma_1k_1k_2(k_1^2-1)}
{\gamma_2k_1(k_2-k_1)+|k_1|\sqrt{\Delta_1}},
\\
\beta_{x\rightarrow y,2}
&=
\frac{\gamma_2k_1(k_2-k_1)-|k_1|\sqrt{\Delta_1}}
{\gamma_1(k_1^2-1)},
&
\beta_{y\rightarrow x,2}
&=
\frac{\gamma_1k_1k_2(k_1^2-1)}
{\gamma_2k_1(k_2-k_1)-|k_1|\sqrt{\Delta_1}},
\end{aligned}
\end{equation}
where $
\Delta_1
=
\gamma_1k_1^2k_2^2-\gamma_1k_1^2-\gamma_1k_2^2+\gamma_1
+\gamma_2^2k_1^2-2\gamma_2^2k_1k_2+\gamma_2^2k_2^2.$
For $j\in\{1,2\}$,
$(\beta_{x\to y,j},\beta_{y\to x,j})$ denotes the $j$th candidate
solution pair. 
This pair is admissible only if $
\sgn(\beta_{x\to y,j})=\sgn(k_1)$
{and} 
$\sgn(\beta_{y\to x,j})=\sgn(k_2).$
\end{proposition}
}

{ When the unobserved confounders have unequal variances and are
correlated, the causal structure becomes more complex, making the
bidirectional causal effects more difficult to distinguish.
Proposition~\ref{prop:UV} provides at most two candidate solution pairs
obtained from a quadratic equation. The existence and number of real
candidate pairs depend on the discriminant $\Delta_1$.Given specified values of the sensitivity parameters, $\Delta_1$ can be evaluated using the observed data, and when $\Delta_1>0$, the quadratic equation yields two distinct real candidate pairs. For each $j\in\{1,2\}$, the candidate pair
$(\beta_{x\to y,j},\beta_{y\to x,j})$ is admissible only if $
\sgn(\beta_{x\to y,j})=\sgn(k_1)$  {and}
$\sgn(\beta_{y\to x,j})=\sgn(k_2)$. 
Thus, the sign restrictions are applied separately to the two candidate
pairs to determine which of them are admissible. If exactly one
candidate pair satisfies these restrictions, the bidirectional causal
effects are uniquely identified. Otherwise, all admissible candidate
pairs constitute the identification results.

In practice, by varying the sensitivity parameters $\gamma_1$ and
$\gamma_2$, we can investigate how changes in the confounding structure
influence the identified bidirectional causal effects. The proof and
further discussion of Proposition~\ref{prop:UV} are provided in
Section~B of the Supplementary Materials.}

Similarly, we can construct plugging-in based estimators and establish the corresponding asymptotic theory by following the approach in Section A.2 of the Supplementary Materials. For simplicity, we omit the details.

\subsection{Further sensitivity analysis for ER assumption}\label{ssec:sensitivity_iv}
{Selecting valid IVs that satisfy Model~\eqref{eq:bidirectional_model} poses significant challenges in practice. A key concern is that the IV $Z$ may have direct effects on $Y^\circ$, and $W$ may have direct effects on $X^\circ$. To address this issue, we extend the sensitivity analysis framework introduced in Section~\ref{ssec:equal-confounding} by incorporating sensitivity parameters for the IVs and evaluating the robustness of our results to possible violations of the exclusion restriction assumption.}

In addition to the sensitivity parameters $\gamma_1$ and $\gamma_2$  in Assumption \ref{assu:U-2}, we introduce two additional sensitivity parameters, $\eta$ and $\delta$, to account for potential violations of the IVs. These parameters represent the extent to which the instrumental variables $Z$ and $W$ exert direct effects on $Y^\circ$ and $X^\circ$, respectively. The modified structural model is given by:
\begin{align}\label{eq:sensitivity-eq}
       \begin{gathered}
X^\circ = \mu_{x0} + \beta_{y \to x} Y^\circ + \mu_{xz} Z + \delta W  + {  U}  , ~~ X =\mathbb{I}\left(X^\circ>0\right),\\
    Y^\circ = \mu_{y0} + \beta_{x \to y} X^\circ +\eta Z+  \mu_{yw}W +   { V}  , ~~ Y =\mathbb{I}\left(Y^\circ>0\right).
\end{gathered}
\end{align}
When the sensitivity parameters $\eta$ and $\delta$  are set to zero, Model \eqref{eq:sensitivity-eq} reduces to Model \eqref{eq:bidirectional_model}.   Given Assumption~\ref{assu:U-2}, we can derive a Probit model similar to Model \eqref{eq:probit_for_assu1}.  {While the coefficients are the same as in the previous model, as they correspond to the regression coefficients in the Probit model, they now involve more complex expressions due to the additional direct effects $\eta$ and $\delta$ introduced in Model \eqref{eq:sensitivity-eq}.} To illustrate these differences in parameter complexity, we present the key parameter expressions below: 
\begin{equation*}
    \begin{aligned}
    \mathrm{pr}(X = 1\mid Z,W) &= \mathrm{pr}(X^\circ  > 0 \mid Z,W)  =  \mathrm{\Phi}(\xi_{x0}+\xi_{xz}Z+\xi_{xw}W),\\  
    \mathrm{pr}(Y = 1\mid Z,W) &= \mathrm{pr}(Y^\circ  > 0 \mid Z,W)  =  \mathrm{\Phi}(\xi_{y0}+\xi_{yz}Z+\xi_{yw}W),
\end{aligned}
\end{equation*}
{where  $\xi_{x0} = \theta_{x0}\big/\sqrt{\lambda_1}$, $\xi_{xz} = \theta_{xz}^*\big/\sqrt{\lambda_1}$, $\xi_{xw} = \theta_{xw}^*\big/\sqrt{\lambda_1}$, $\theta_{xz}^* = c(\mu_{xz}+\beta_{y\rightarrow x}\eta) $, $\theta_{xw}^* = c(\delta+\beta_{y\rightarrow x}\mu_{yw}) $, $\theta_{yz}^* = c(\beta_{x\rightarrow y}\mu_{xz}+\eta)$, and $\theta_{yw}^* = c(\beta_{x\rightarrow y}\delta+\mu_{yw})$ and similarly for $\xi_{y0}$, $\xi_{yz}$, and $\xi_{yw}$.} The coefficients 
  $c$, $\lambda_1$, $\lambda_2$, $ \theta_{x0}$,  $ \theta_{y0}$,  $\theta_{xu}  $, $\theta_{xv} $, $\theta_{yu} $, and $\theta_{yv} $ are already defined in~\eqref{eq:identify_start} and~\eqref{eq:probit_for_assu1}.

Given the sensitivity parameters $\eta$ and $\delta$, to establish identifiable results, we introduce reparameterizations which will be used in the following theoretical results. We define $\eta_0 = \eta / \mu_{xz}$, which represents the relative strength of the effect of $Z$ on $(X^\circ,Y^\circ)$, and $\delta_0 = \delta / \mu_{yw}$, which represents the relative strength of the effect of $W$ on $(X^\circ,Y^\circ)$.

Building on Proposition~\ref{prop:UV}, we further extend our identification strategy to enable more comprehensive sensitivity analysis by incorporating direct effect parameters ($\eta_0$, $\delta_0$) under Model~\eqref{eq:sensitivity-eq} ({see Proposition~S2 in Section C.1 of Supplementary Materials}).

However, solving the quartic equation  in Proposition S2 poses significant computational challenges in practical applications. Intuitively, this difficulty arises because the direct effects of $Z$ and $W$ on $X^\circ$ and $Y^\circ$ are inherently intertwined, making it challenging to disentangle their individual contributions to the causal effects. To make the sensitivity analysis more tractable and practically interpretable, we examine three special cases that yield simplified analytical expressions, thereby facilitating implementation in empirical applications. 
Figure~\ref{fig:iv figure} illustrates the causal frameworks under the sensitivity analysis settings specified in the following three corollaries. 

\begin{figure}[t]
\centering
\captionsetup{font=small}

\begin{subfigure}[t]{0.31\textwidth}
\centering
\resizebox{\linewidth}{!}{%
\begin{tikzpicture}[>=latex, scale=0.65]
\tikzstyle{node} = [draw, circle, thick, 
                    minimum size=1cm,
                    inner sep=1pt,
                    text width=0.8cm, 
                    align=center]
\node[node] (Z) at (-3.3,0) {$Z$};
\node[node] (X_) at (0,0) {$X^\circ$};
\node[node] (X) at (0,-2.8) {$X$};
\node[node] (Y_) at (3.5,0) {$Y^\circ$};
\node[node] (Y) at (3.5,-2.8) {$Y$};
\node[node, fill=gray!30] (U) at (0,2.75) {$U$}; 
\node[node, fill=gray!30] (V) at (3.5,2.75) {$V$};
\node[node] (W) at (6.8,0) {$W$};
\draw[thick, ->] (Z) -- (X_);
\draw[thick, ->] (X_) -- (X);
\draw[thick, ->] (Y_) -- (Y);
\draw[thick, ->] (X_) to[out=20,in=160] (Y_);
\draw[thick, ->] (Y_) to[out=200,in=340] (X_);
\draw[thick, ->] (U) -- (X_);
\draw[thick, ->] (V) -- (Y_);
\draw[thick, ->] (W) -- (Y_);
\draw[thick, ->] (Z) to[out=340,in=220] (Y_);
\end{tikzpicture}}
\caption{ }
\label{fig:excl-viol-wx}
\end{subfigure}
\hfill
\begin{subfigure}[t]{0.31\textwidth}
\centering
\resizebox{\linewidth}{!}{%
\begin{tikzpicture}[>=latex, scale=0.65]
\tikzstyle{node} = [draw, circle, thick, 
                    minimum size=1cm,
                    inner sep=1pt,
                    text width=0.8cm, 
                    align=center]
\node[node] (Z) at (-3.3,0) {$Z$};
\node[node] (X_) at (0,0) {$X^\circ$};
\node[node] (X) at (0,-2.8) {$X$};
\node[node] (Y_) at (3.5,0) {$Y^\circ$};
\node[node] (Y) at (3.5,-2.8) {$Y$};
\node[node, fill=gray!30] (U) at (0,2.75) {$U$}; 
\node[node, fill=gray!30] (V) at (3.5,2.75) {$V$};
\node[node] (W) at (6.8,0) {$W$};
\draw[thick, ->] (Z) -- (X_);
\draw[thick, ->] (X_) -- (X);
\draw[thick, ->] (Y_) -- (Y);
\draw[thick, ->] (X_) to[out=20,in=160] (Y_);
\draw[thick, ->] (Y_) to[out=200,in=340] (X_);
\draw[thick, ->] (U) -- (X_);
\draw[thick, ->] (V) -- (Y_);
\draw[thick, ->] (W) -- (Y_);
\draw[thick, ->] (W) to[out=200,in=320] (X_); 
\end{tikzpicture}}
\caption{ }
\label{fig:excl-viol-zy}
\end{subfigure}
\hfill
\begin{subfigure}[t]{0.31\textwidth}
\centering
\resizebox{\linewidth}{!}{%
\begin{tikzpicture}[>=latex, scale=0.65]
\tikzstyle{node} = [draw, circle, thick, 
                    minimum size=1cm,
                    inner sep=1pt,
                    text width=0.8cm, 
                    align=center]
\node[node] (Z) at (-3.3,0) {$Z$};
\node[node] (X_) at (0,0) {$X^\circ$};
\node[node] (X) at (0,-2.8) {$X$};
\node[node] (Y_) at (3.5,0) {$Y^\circ$};
\node[node] (Y) at (3.5,-2.8) {$Y$};
\node[node, fill=gray!30] (U) at (0,2.75) {$U$}; 
\node[node, fill=gray!30] (V) at (3.5,2.75) {$V$}; 
\node[node] (W) at (6.8,0) {$W$};
\draw[thick, ->] (Z) -- (X_);
\draw[thick, ->] (X_) -- (X);
\draw[thick, ->] (Y_) -- (Y);
\draw[thick, ->] (X_) to[out=20,in=160] (Y_);
\draw[thick, ->] (Y_) to[out=200,in=340] (X_);
\draw[thick, ->] (U) -- (X_);
\draw[thick, ->] (V) -- (Y_);
\draw[thick, ->] (W) -- (Y_);
\draw[thick, -] (U) -- (V) node[midway, above] {{ $\gamma_{2}=1$}};
\draw[thick, ->] (W) to[out=200,in=320] (X_);
\draw[thick, ->] (Z) to[out=340,in=220] (Y_);
\end{tikzpicture}}
\caption{  }
\label{fig:excl-viol-both-edges}
\end{subfigure}
\caption{Three graphical illustrations of Model \eqref{eq:sensitivity-eq}.}
\label{fig:iv figure}
\end{figure}
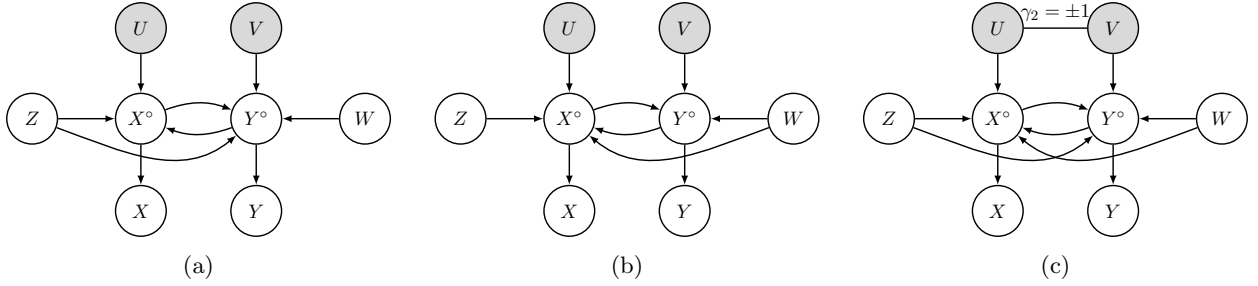
\begin{corollary}\label{corollary:1} { Under Assumption~\ref{assu:U-2}, with the settings $\delta_0 = 0$, $\gamma_1 = 1$, $\gamma_2 = 0$, and an arbitrary $\eta_0$ (illustrated in Figure~\ref{fig:excl-viol-wx}), the bidirectional causal effects can be obtained through the following two candidate solution pairs:
\begin{equation*}
    \begin{aligned}
    &\beta_{y\rightarrow x ,1} = \frac{-s_2+ \sqrt{s_2^2-4s_1s_3}}{2s_1}, \ \ \beta_{x\rightarrow y,1} = \eta_0(t_1t_2-1)+\frac{2s_1t_1t_2}{-s_2+ \sqrt{s_2^2-4s_1s_3}},\\
    &\beta_{y\rightarrow x ,2} = \frac{-s_2- \sqrt{s_2^2-4s_1s_3}}{2s_1}, \ \ \beta_{x\rightarrow y,2} = \eta_0(t_1t_2-1)+\frac{2s_1t_1t_2}{-s_2- \sqrt{s_2^2-4s_1s_3}},
\end{aligned}
\end{equation*}
 where $t_1 =   {\xi_{xw}}/ {\xi_{xz}} $, $t_2 = {\xi_{yz}}/{\xi_{yw}}$, 
        $ t_3 = {\xi_{xz}}/{\xi_{yz}}$,  $t_4 = {\xi_{xw}}/{\xi_{yw}} $, $s_1 =\eta_0^2(t_1{t_2}-1)^2 -t_4^2+1$, $s_2 = 2t_1t_2\eta_0(t_1t_2-1)$, and $s_3 =  t_1^2t_2^2 - t_4^2$. {For each $j\in\{1,2\}$,
$(\beta_{x\to y,j},\beta_{y\to x,j})$ denotes the $j$th candidate
solution pair. The $j$th candidate pair is admissible only if $\sgn(\beta_{y\to x,j})=\sgn(t_4)$.} } \end{corollary}

{\begin{corollary}\label{corollary:2} { Under Assumption~\ref{assu:U-2}, with the settings $\eta_0 = 0$, $\gamma_1 = 1$, $\gamma_2 = 0$, and an arbitrary $\delta_0$ (illustrated in Figure~\ref{fig:excl-viol-zy}), the bidirectional causal effects can be obtained through the following two candidate solution pairs:
\begin{equation*}
    \begin{aligned}
    &{ \beta_{x\rightarrow y,1}} = \frac{-s_5+ \sqrt{s_5^2-4s_4s_6}}{2s_4},\ \ {\beta_{y\rightarrow x,1}} = \delta_0(t_1t_2-1)+\frac{2s_4t_1t_2}{-s_5+ \sqrt{s_5^2-4s_4s_6}},\\
    &{\beta_{x\rightarrow y,2}} = \frac{-s_5- \sqrt{s_5^2-4s_4s_6}}{2s_4},\ \ { \beta_{y\rightarrow x,2}} = \delta_0(t_1t_2-1)+\frac{2s_4t_1t_2}{-s_5- \sqrt{s_5^2-4s_4s_6}},
\end{aligned}
\end{equation*}
 where $t_1   $, $t_2 $,    
        $ t_3  $ and   $t_4 $ are already defined in Corollary \ref{corollary:1}, $s_4 =t_3t_4+t_3t_4\delta_0^2(t_1{t_2}-1)^2 -t_1t_2$, $s_5 = 2t_4^2\delta_0(t_1t_2-1)$, and $s_6 =  t_1t_2(t_4^2 - 1)$. { For each $j\in\{1,2\}$,
$(\beta_{x\to y,j},\beta_{y\to x,j})$ denotes the $j$th candidate
solution pair. The $j$th candidate pair is admissible only if $\sgn(\beta_{x\to y,j})=\sgn(t_3)$.} } \end{corollary}}

{Corollary~\ref{corollary:1} examines the identification of the
bidirectional causal effects when $Z$ has a direct effect on
$Y^\circ$, under the assumptions that $W$ is a valid instrumental
variable and that the unobserved confounders $U$ and $V$ are
independent. Corollary~\ref{corollary:2} considers the corresponding
setting in which $W$ has a direct effect on $X^\circ$. 
Each corollary yields at most two candidate solution pairs. In
Corollary~\ref{corollary:1}, a candidate pair indexed by $j$ is
admissible only if $
\sgn(\beta_{y\to x,j})=\sgn(t_4),$ 
whereas in Corollary~\ref{corollary:2}, it is admissible only if $
\sgn(\beta_{x\to y,j})=\sgn(t_3).$ 
If exactly one candidate pair satisfies the   sign
restriction, the bidirectional causal effects are uniquely identified.
This approach is motivated by sensitivity analyses for violations of
the exclusion restriction in traditional IV 
frameworks \citep{Angrist1996}.}

Beyond the scenarios in Corollaries \ref{corollary:1} and \ref{corollary:2}, we explore cases where the unobserved confounders $U$ and $V$ are correlated, while simultaneously allowing both $Z$ and $W$ to have direct effects on $X^\circ$ and $Y^\circ$.

{\begin{corollary}\label{corollary:3}
We introduce the signal-to-noise ratios $\eta_0 = \eta / \sigma$ and $\delta_0 = \delta / \sigma$ to reparameterize $\eta$ and $\delta$ in Model~\eqref{eq:sensitivity-eq}. We then consider the setting with $\gamma_1=1$, { $\gamma_2=1$}, and {fixed values of $\delta_0$ and $\eta_0$} (as illustrated in Figure~\ref{fig:excl-viol-both-edges}). { Suppose that both $\beta_{x\to y} > -1$ and $\beta_{y\to x} > -1$, that $\beta_{x\to y}\beta_{y\to x}\neq 1$.} { If $\beta_{x\to y}\beta_{y\to x}<1$,} the identification expressions for the bidirectional causal effects are given by:
\begin{equation}
\label{solution-case3}
\begin{aligned}
&\beta_{y \to x} = \frac{( \xi_{yz} - \xi_{xz})(\xi_{xw}-\delta_0)}{(\xi_{xw} - \xi_{yw})(\xi_{xz}-\eta_0)},~~~
\beta_{x \to y} = \frac{(\xi_{xw} - \xi_{yw})(\xi_{yz}-\eta_0)}{(\xi_{yz} - \xi_{xz})(\xi_{yw}-\delta_0)}.
\end{aligned}
\end{equation}
{ If $\beta_{x\to y}\beta_{y\to x}>1$,} the identification expressions for the bidirectional causal effects are given by:
\begin{equation}
\label{solution-case3-2}
\begin{aligned}
&\beta_{y \to x} = \frac{( \xi_{yz} - \xi_{xz})(\xi_{xw}+\delta_0)}{(\xi_{xw} - \xi_{yw})(\xi_{xz}+\eta_0)},~~~
\beta_{x \to y} = \frac{(\xi_{xw} - \xi_{yw})(\xi_{yz}+\eta_0)}{(\xi_{yz} - \xi_{xz})(\xi_{yw}+\delta_0)}.
\end{aligned}
\end{equation}
\end{corollary}}

{Equations~\eqref{solution-case3} and~\eqref{solution-case3-2} each yield one candidate pair. They provide identification expressions under different ranges of the causal effect parameters. In fact, except for values that make the denominators undefined, namely \(\beta_{y\to x}=-1\), \(\beta_{x\to y}=-1\), and \(\beta_{y\to x}\beta_{x\to y}=1\), identification expressions exist for all other parameter regions. More detailed results are presented in Section~C.4 of the Supplementary Materials. Although multiple pairs may arise during the identification process due to the sign configurations of the parameters, under each sign configuration the identification expression is unique and has a neat closed-form expression; these solutions together constitute the complete set of candidate pairs. Therefore, with further prior information, for different signal-to-noise ratios, we can compute the causal effects based on the candidate pairs and subsequently perform selection to achieve more accurate identification. Figure S1 of the Supplementary Materials illustrates the regions of these configurations.}

{Sections C.2--C.4 of the Supplementary Materials provide the detailed proofs of Corollaries \ref{corollary:1}--\ref{corollary:3}, while Section~C.5 presents the discussion of Corollary~\ref{corollary:3} under the alternative parameterization based on $\eta/\mu_{xz}$ and $\delta/\mu_{yw}$.}

\section{Numerical Experiments}\label{section:simulation}

\subsection{Simulation studies}\label{subsection:simulation_settings}

In this section, we conduct numerical simulations to evaluate the finite sample performance of the proposed method in Section \ref{section:methodology}.   We generate observed covariate $Q \sim N(0, 1)$ and unobserved confounders $U, V \sim N(0, 0.75^2)$. Instrumental variables $Z$ and $W$ follow either $N(0, 1)$ (Scenario 1) or $\text{Unif}(-1, 1)$ (Scenario 2). Latent variables $X^\circ$ and $Y^\circ$ are generated from Model~\eqref{eq:bidirectional_model} with parameters $\mu_{x0} =\mu_{y0} = 0$, $\beta_{y \to x} = 0.45$, $\beta_{x \to y} = -0.25$, $\mu_{xz} = \mu_{yw} = 0.65$, $\mu_{xq} =\mu_{yq}= 0.15$, and then thresholded to obtain binary $X$ and $Y$.

\begin{figure}[htbp]
    \centering
    \begin{subfigure}[t]{0.49\textwidth}
    \includegraphics[width=\linewidth, height = 4.8cm]{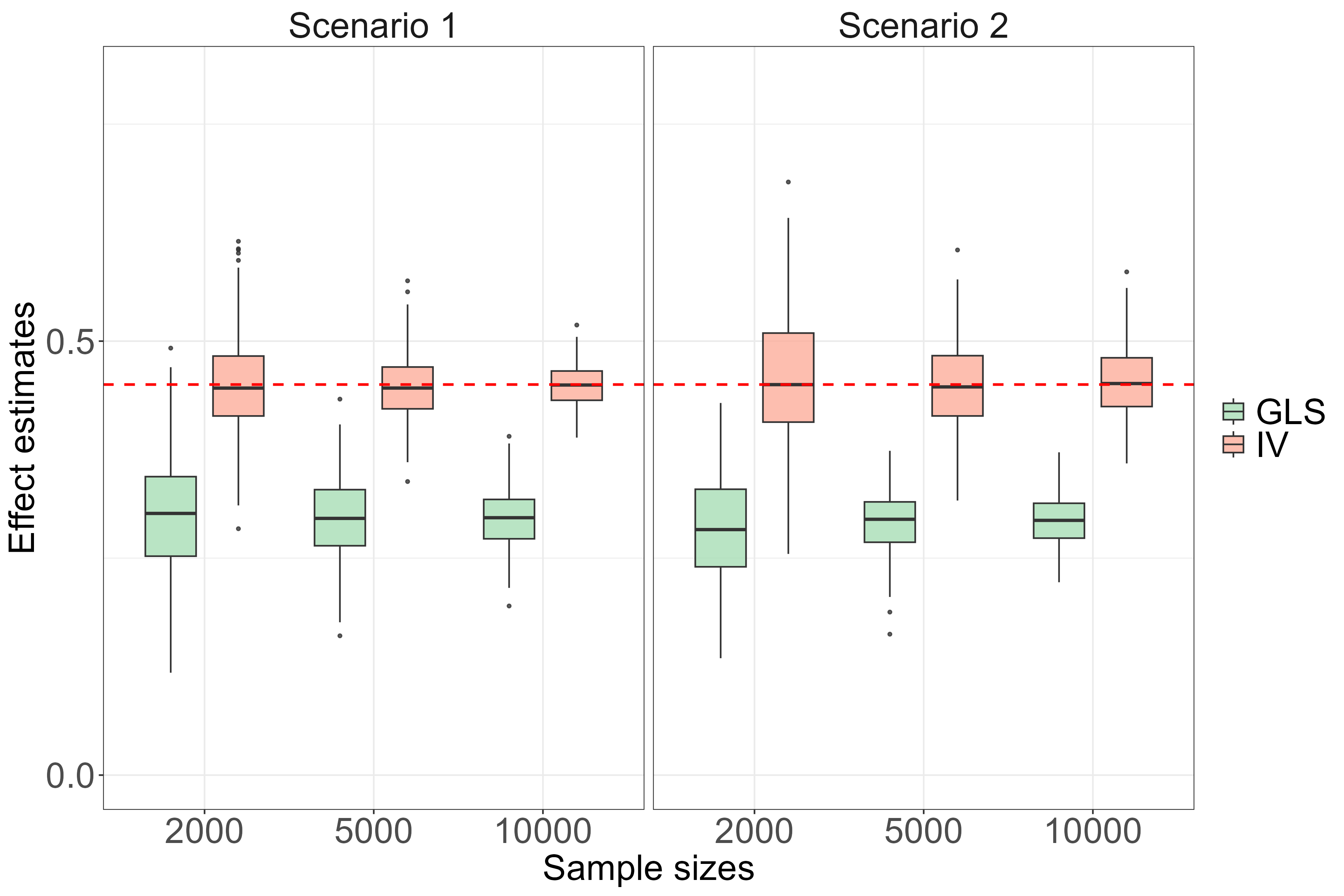}
        \caption{Simulation results of $\beta_{x\to y}$.}
        \label{fig:p_x2y_simulation}
    \end{subfigure}
    \hfill 
    \begin{subfigure}[t]{0.49\textwidth}
    \includegraphics[width=\linewidth, height = 4.8cm]{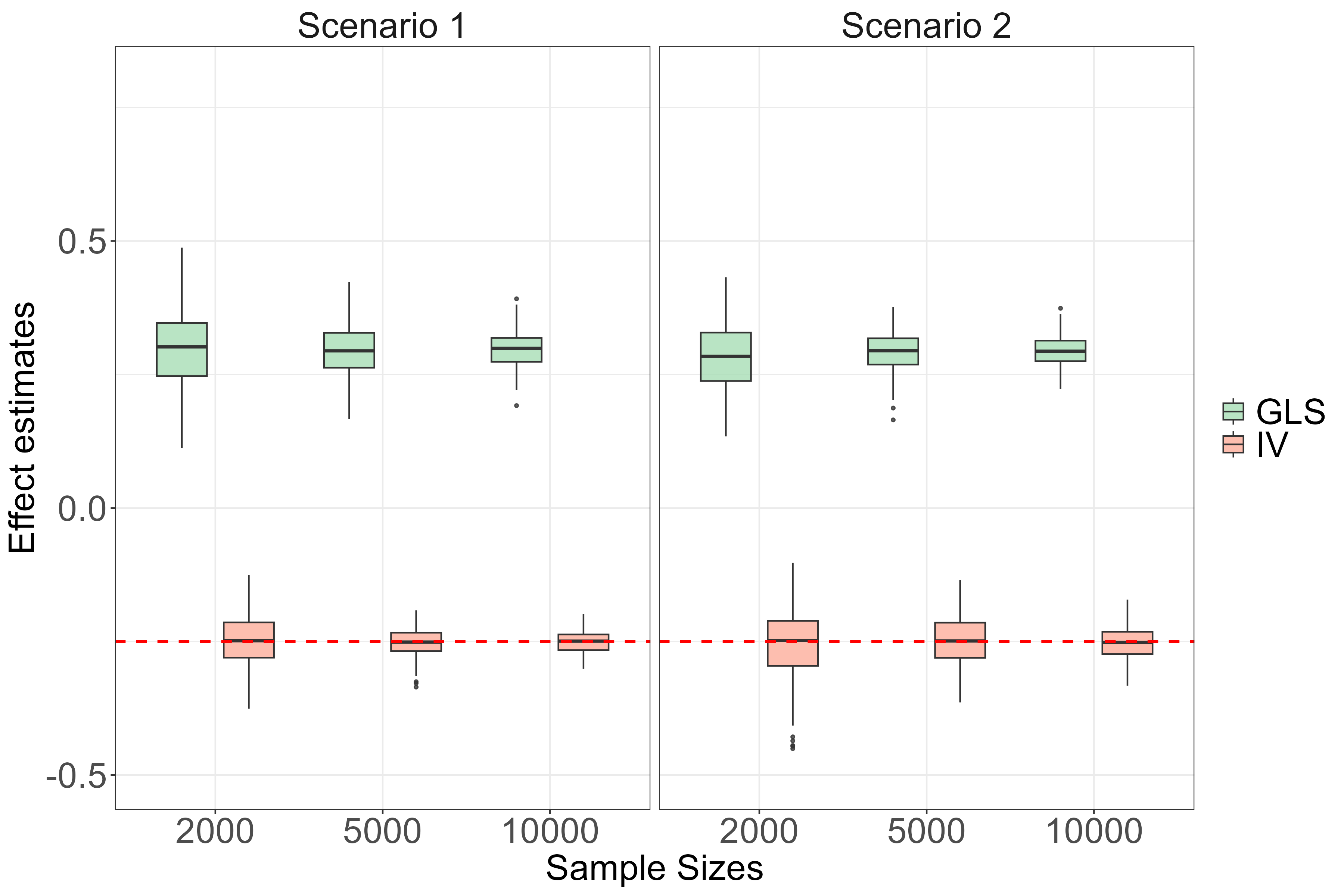}
        \caption{Simulation results of $\beta_{y \to x}$.} 
        \label{fig:p_y2x_simulation}
    \end{subfigure}
    \caption{Boxplot of simulation results under bidirectional framework.} 
    \label{fig:causal_simulations}
\end{figure}

For the two data-generating scenarios, we consider two estimation methods with sample sizes $N = 2000, 5000, $ and $ 10000$.  {The first method (GLS) is the direct regression approach, where we fit two Probit models: one by regressing $Y$ on $(X,Z,W,Q)$, and another by regressing $X$ on $(Y,Z,W,Q)$,} the resulting coefficients on $X$ and $Y$ are then interpreted as estimates of $\hat{\beta}_{x \to y}$ and $\hat{\beta}_{y \to x}$, respectively.  The second method (IV) is Algorithm S1 in Section A.2 of the Supplementary Materials. We perform Probit regression on $X$, $Y$, $Z$, $W$ and $Q$ to estimate  the coefficients in Equation~\eqref{eq:probit_for_assu1},  namely, { $\xi_{xz}$,  $\xi_{xw}$,  $\xi_{yz}$ and $\xi_{yw}$}. These estimated coefficients are then substituted into the identification formula~\eqref{eq:beta_identify} to obtain the estimates $\hat{\beta}_{x \to y}$ and $\hat{\beta}_{y \to x}$.

The estimation results show that the proposed method performs well across both scenarios,  yielding consistent estimates. In contrast,  the direct regression approach produces estimates that deviate significantly from the true values in both scenarios. Furthermore,  we compute the standard deviation and bias of the proposed estimators over 200 repeated simulations. Figure~\ref{fig:causal_simulations} presents the simulation results for the bidirectional causal effect, where the red represents our proposed method and the green represents the naive regression method. The corresponding numerical results are summarized in Table~\ref{tab:simulation_sd_bias}. It can be observed that the standard deviation and bias of the estimators decrease as the sample size increases across both scenarios.


\begin{table}[h]
  \centering
  \caption{{ Standard deviation and bias of estimators across scenarios and methods. All values are multiplied by $10^4$. Detailed results are provided in Section D.1 of the Supplementary Materials.}}
  \label{tab:simulation_sd_bias}
 \small
 \resizebox{0.995\textwidth}{!}
 {
\begin{tabular}{clccccccccccc}
\toprule
                                         &  &             &  & \multicolumn{2}{c}{Scenario 1 (GLS)} & \multicolumn{2}{c}{Scenario 2 (GLS)} &  & \multicolumn{2}{c}{Scenario 1 (IV)} & \multicolumn{2}{c}{Scenario 2 (IV)} \\ \cline{5-8} \cline{10-13} 
                                         &  & $N$ &  & 2000              & 10000            & 2000              & 10000            &  & 2000             & 10000            & 2000             & 10000            \\   \cline{3-3} \cline{5-8} \cline{10-13} 
\multirow{2}{*}{$\hat{\beta}_{x \to y}$} &  & bias        &  & -1510             & -1540            & -1640             & -1570            &  & -2               & 10               & 80               & -8               \\
                                         &  & sd          &  & 724               & 358              & 635               & 288              &  & 535              & 242              & 941              & 407              \\
\multirow{2}{*}{$\hat{\beta}_{y \to x}$} &  & bias        &  & 5490              & 5460             & 5370              & 5440             &  & -45              & -7               & 10               & -31              \\
                                         &  & sd          &  & 718               & 355              & 637               & 289              &  & 500              & 208              & 755              & 354              \\ \toprule
\end{tabular}
}
\end{table}

\subsection{Sensitivity analysis for Assumption~\ref{assu:U}}\label{ssec:UV_simulation}
In this section, we conduct a sensitivity analysis of Assumption~\ref{assu:U}, focusing on the influence of both the variance relationship and covariance of the unobserved confounders $U$ and $V$ on the estimation results. Accordingly, we modify the second step of the data generation process in Section~\ref{subsection:simulation_settings} as follows: Under Assumption~\ref{assu:U-2}, we generate the unobserved confounders $U$ and $V$, where $U \sim N(0, \gamma_1 \sigma^2)$, $V \sim N(0, \sigma^2)$, and $\text{cov}(U, V) = \gamma_2 \sigma^2$, with $\sigma = 0.75$. The parameter $\gamma_1$ varies within the range (0.1, 1) at intervals of 0.02, while $\gamma_2$ varies within the range (-0.6, 0.6) at intervals of 0.02. All other settings remain consistent with Section~\ref{subsection:simulation_settings}. 

\begin{figure}[htbp]
    \centering
    \includegraphics[width=0.995\linewidth]{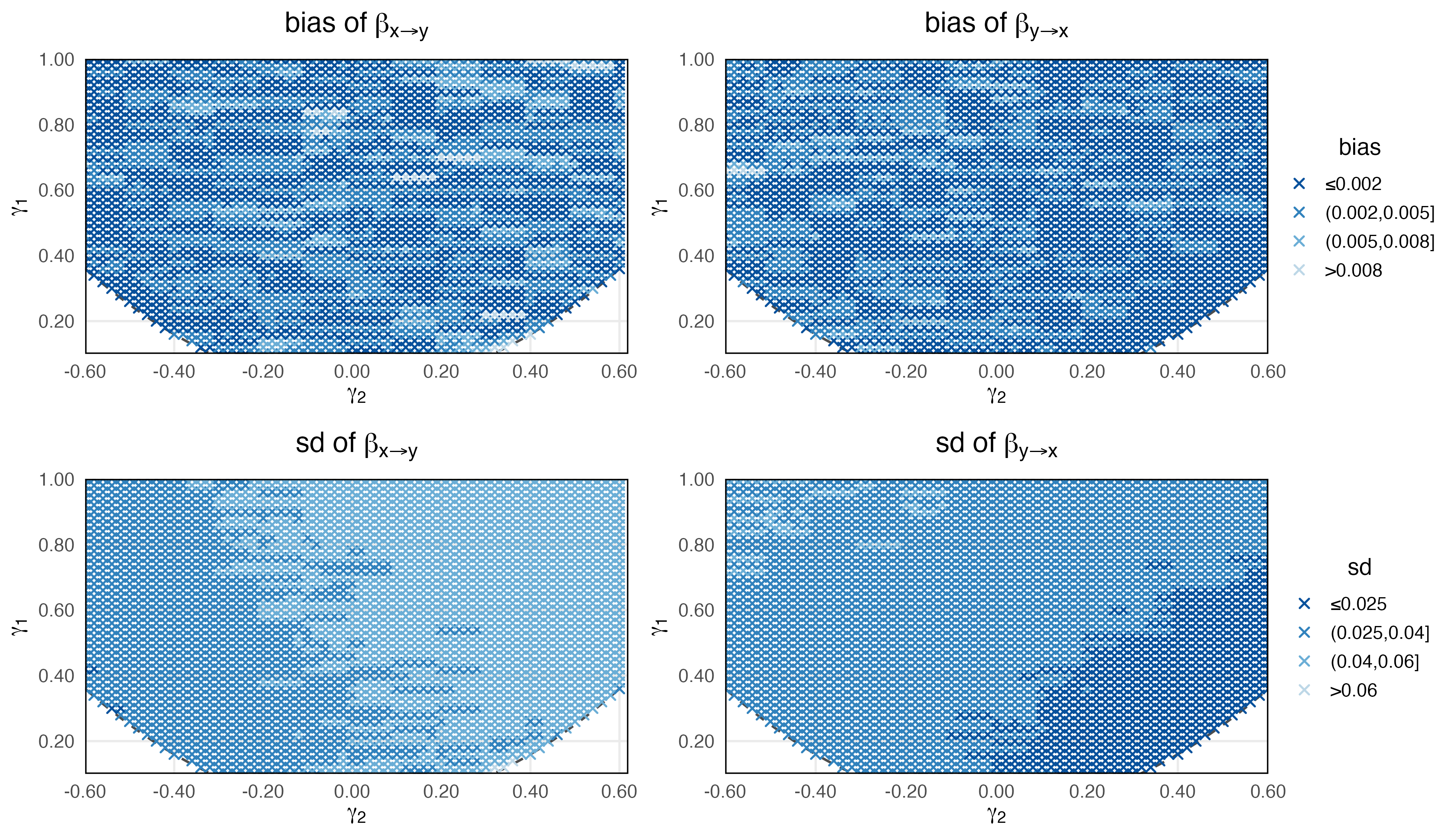}
\caption{Bias and standard deviation estimation results.}
    \label{fig:bias_combind}
\end{figure}

{ For each parameter combination $(\gamma_1,\gamma_2)$ and each of
the 200 simulated datasets, we calculate the two candidate solution
pairs given in Proposition~\ref{prop:UV} and retain those satisfying
the sign restrictions
$\sgn(\beta_{x\to y,j})=\sgn(k_1)$ 
 {and} 
$\sgn(\beta_{y\to x,j})=\sgn(k_2).$
The estimates reported below refer to the resulting admissible
candidate pairs.}
Figure~\ref{fig:bias_combind} presents the bias and standard deviation of   causal effect estimates across different sensitivity parameter values for a sample size of $N=10000$. The black curve corresponds to the truncation of the parabola defined by $\gamma_1 = \gamma_2^2$. As stated in Proposition~\ref{prop:UV}, within the parabola, specific point estimates of the causal effects can be obtained once the two sensitivity parameter values are determined.

The first row of Figure~\ref{fig:bias_combind} presents the bias of the bidirectional causal effect estimates $\beta_{x \to y}$ and $\beta_{y \to x}$ across different combinations of sensitivity parameters $(\gamma_1, \gamma_2)$. {The gradient from light to dark corresponds to values from large to small. Across a broad sensitivity range,  the estimation bias   remains small, around 0.002 (mainly shown in dark colors), with only minor fluctuations. The second row displays the standard deviations of the estimates, with the color gradient again corresponding to their values. Although the standard deviations vary across parameter values, they remain consistently below 0.06, indicating the robustness of the method.}

\subsection{Sensitivity analysis simulation for ER assumption}\label{subsection:sensitivity analysis simulation}

This section conducts numerical simulations for the three instrumental variable sensitivity analysis settings discussed in Section~\ref{ssec:sensitivity_iv}. We assess how violating Model~\eqref{eq:bidirectional_model} by introducing direct effects from instrumental variables to target variables impacts causal effect estimation under finite samples. The simulation setup follows Section~\ref{subsection:simulation_settings}, with modifications corresponding to Corollaries~\ref{corollary:1}--\ref{corollary:3}. We consider three cases: Case 1 (Corollary \ref{corollary:1}) varies $\eta_0 \in [-0.16, 0.16]$ with $\delta_0= 0$, $\gamma_1 = 1$, $\gamma_2 = 0$; Case 2 (Corollary \ref{corollary:2}) varies $\delta_0 \in [-0.16, 0.16]$ with $\eta_0= 0$, $\gamma_1 = 1$, $\gamma_2 = 0$; Case 3 (Corollary \ref{corollary:3}) varies $\eta_0 = \delta_0 \in [-0.16, 0.16]$ with $\gamma_1 = \gamma_2 = 1$. All sensitivity parameters vary by increments of $0.02$.

{ For Cases~1 and~2, the two candidate solution pairs are
calculated for each value of the sensitivity parameter. In Case~1, we
retain the candidate pairs satisfying $
\sgn(\beta_{y\to x,j})=\sgn(t_4),$ 
whereas in Case~2, we retain the candidate pairs satisfying $
\sgn(\beta_{x\to y,j})=\sgn(t_3).$ 
The estimates reported below correspond to the resulting admissible
candidate pairs.}

\begin{figure}[htbp]
    \centering
\includegraphics[width=0.995\linewidth]{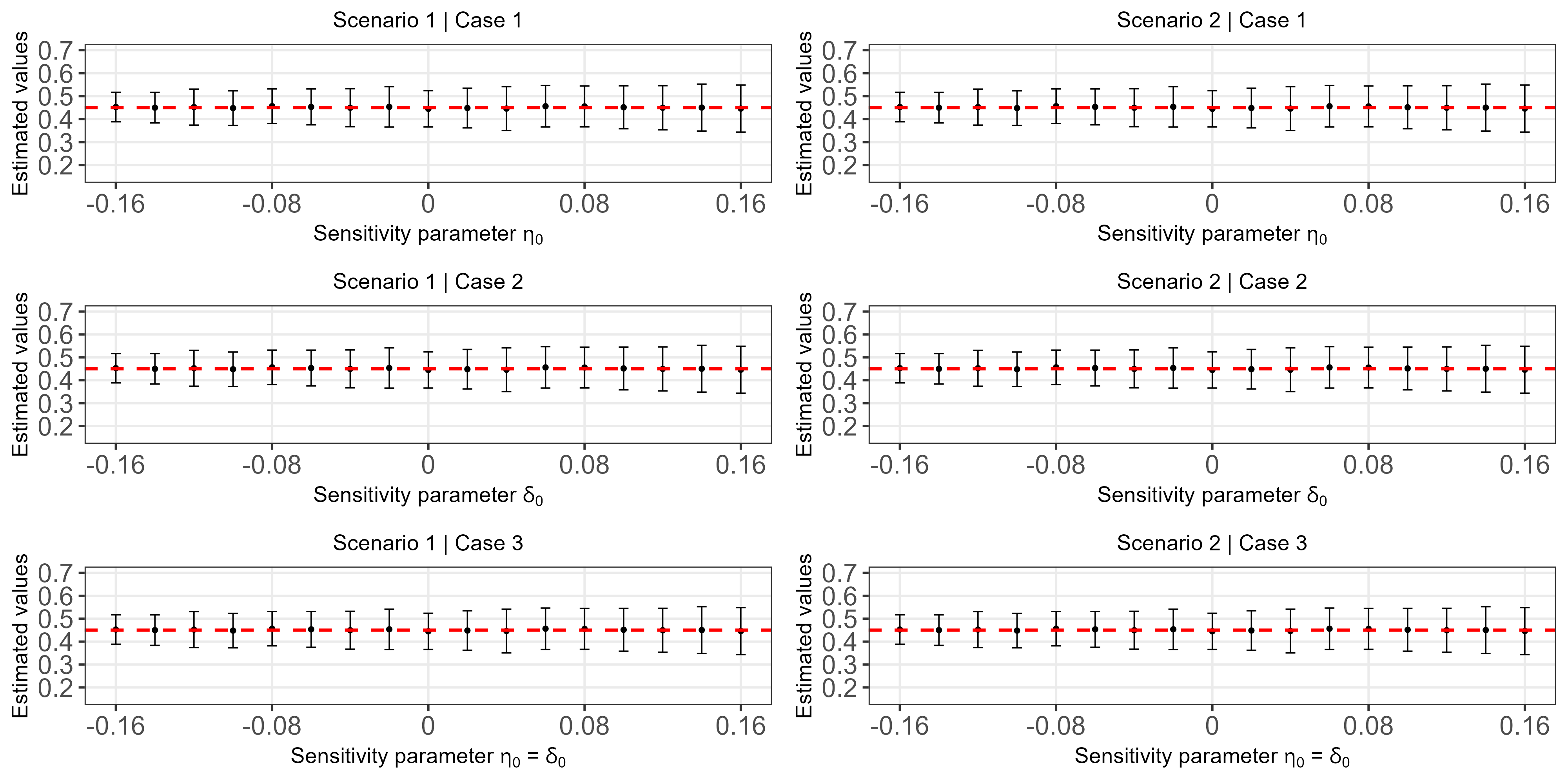}
\caption{The point estimate and 95\% confidence interval results of $\beta_{x \rightarrow y}$.}
    \label{fig:errorbar3case_beta_x2y}
\end{figure}

Based on the simulation results from 200  repeated simulations under different combinations of parameters and data-generating scenarios, we plot  error bar graphs in  Figures~\ref{fig:errorbar3case_beta_x2y}.  
The red dashed lines  represent the given true value of the causal effect parameter. It can be observed that across different parameter values, the point estimates under both data-generating scenarios remain consistently stable around the true value. Furthermore, when the arbitrary sensitivity parameters in the first two settings are changed to zero, the estimation results are equivalent to those under the specification of Model~\eqref{eq:bidirectional_model}. Simultaneously, as the sample size increases, the estimation error decreases accordingly. These findings demonstrate that the proposed method retains its robustness even when the instrumental variables $Z$ and $W$ violate the assumptions of Model~\eqref{eq:bidirectional_model} to some extent.  Here, we present the results for $\beta_{x\to y}$ under three settings with a sample size of $N=10000$. Simulation results for other sample sizes and for $\beta_{y\to x}$ are provided in detail in {Section D.2 of the Supplementary Materials.}

\section{Application}\label{section:case}
\subsection{Description}

According to the Centers for Disease Control and Prevention (CDC), heart disease is the leading cause of death among most racial groups in the United States, including African Americans, American Indians/Alaska Natives, and Whites. In addition, nearly half (47\%) of US adults have at least one of the three major risk factors: high blood pressure, high cholesterol, or smoking. Other important risk factors include diabetes, obesity (high body mass index, BMI), physical inactivity, and excessive alcohol consumption.

In this study, we apply our method to investigate the bidirectional causal relationship between heart disease and diabetes. The data come from the 2023 Behavioral Risk Factor Surveillance System (BRFSS) by the CDC, containing 319795 records and 18 variables including coronary heart disease or myocardial infarction, diabetes status, smoking, BMI, and other health factors. We focus on the bidirectional relationship between $X$ (coronary heart disease or myocardial infarction) and $Y$ (diabetes). The remaining 16 variables (e.g., age, gender, race) are in Table S2 of Section D.3 in the Supplementary Materials. We standardize BMI prior to analysis given its larger scale compared to binary variables.

Since stroke is clinically associated with heart disease but has limited impact on diabetes, and higher BMI increases diabetes risk while having limited effect on heart disease, {we treat ``stroke'' as the instrumental variable $Z$ for $X^\circ$ and ``BMI'' as the instrumental variable $W$ for $Y^\circ$.}

\subsection{Analysis under the  assumptions in Section \ref{ssec:Identification}}\label{ssec:case estimate}

For the heart disease dataset, we apply our proposed method under Assumption~\ref{assu:U} to estimate the bidirectional causal effects between heart disease and diabetes, while also providing results from the direct regression approach for comparison. Based on 200 bootstrap replications, we compute the standard deviations (SD) and 95\% confidence intervals (CIs) for $\beta_{x \to y}$ and $\beta_{y \to x}$. All results are presented in Table~\ref{tab:bootstrap_results}.

\begin{table}[htbp]
  \centering
 \caption{Estimation results of the heart disease application.}
  \label{tab:bootstrap_results}
  \footnotesize
  \resizebox{0.998\textwidth}{!}
  {
\begin{tabular}{ccccccccc}
\toprule
                        &  & Estimate & SD     & 95\% CI                      &                   & Estimate & SD     & 95\% CI                      \\ \cline{3-5} \cline{7-9} 
                        &  & \multicolumn{3}{c}{GLS estimation}                          &                   & \multicolumn{3}{c}{IV estimation}                           \\
$\hat{\beta}_{x \to y}$ &  & 0.2510   & 0.0097 & [0.2321, 0.2698] & \multirow{2}{*}{} & 0.3108   & 0.0262 & [0.2594, 0.3621] \\
$\hat{\beta}_{y \to x}$ &  & 0.2554   & 0.0090 & [0.2379, 0.2729] &                   & 0.1786   & 0.0154 & [0.1484, 0.2088] \\ \toprule
\end{tabular}}
\end{table}

We find that both point estimates are positive,  indicating a bidirectional positive association between $X$ and $Y$, that is,  coronary heart disease or myocardial infarction may have a significant positive effect on diabetes,  and vice versa. The 95\% confidence intervals obtained via bootstrap do not include zero,  further supporting the presence of statistically significant positive effects in both directions. These findings imply that the onset of heart disease may elevate the risk of developing diabetes, potentially due to physiological stress, inflammation, or lifestyle changes following a cardiovascular event. Conversely, individuals with diabetes may face an increased risk of heart disease, {consistent with existing clinical evidence that links poor glycemic control to vascular damage and atherosclerosis  \citep{Brownlee2001}.} In addition, the direct regression approach yields point estimates with the same direction of effect, suggesting that the estimation results obtained using our proposed method are robust and reliable.

\subsection{{Sensitivity analysis for Assumption~\ref{assu:U}}}\label{ssec:case for UV}
\begin{figure}[htbp]
    \centering
    \begin{subfigure}{0.49\linewidth}
        \centering
        \includegraphics[width=\linewidth]{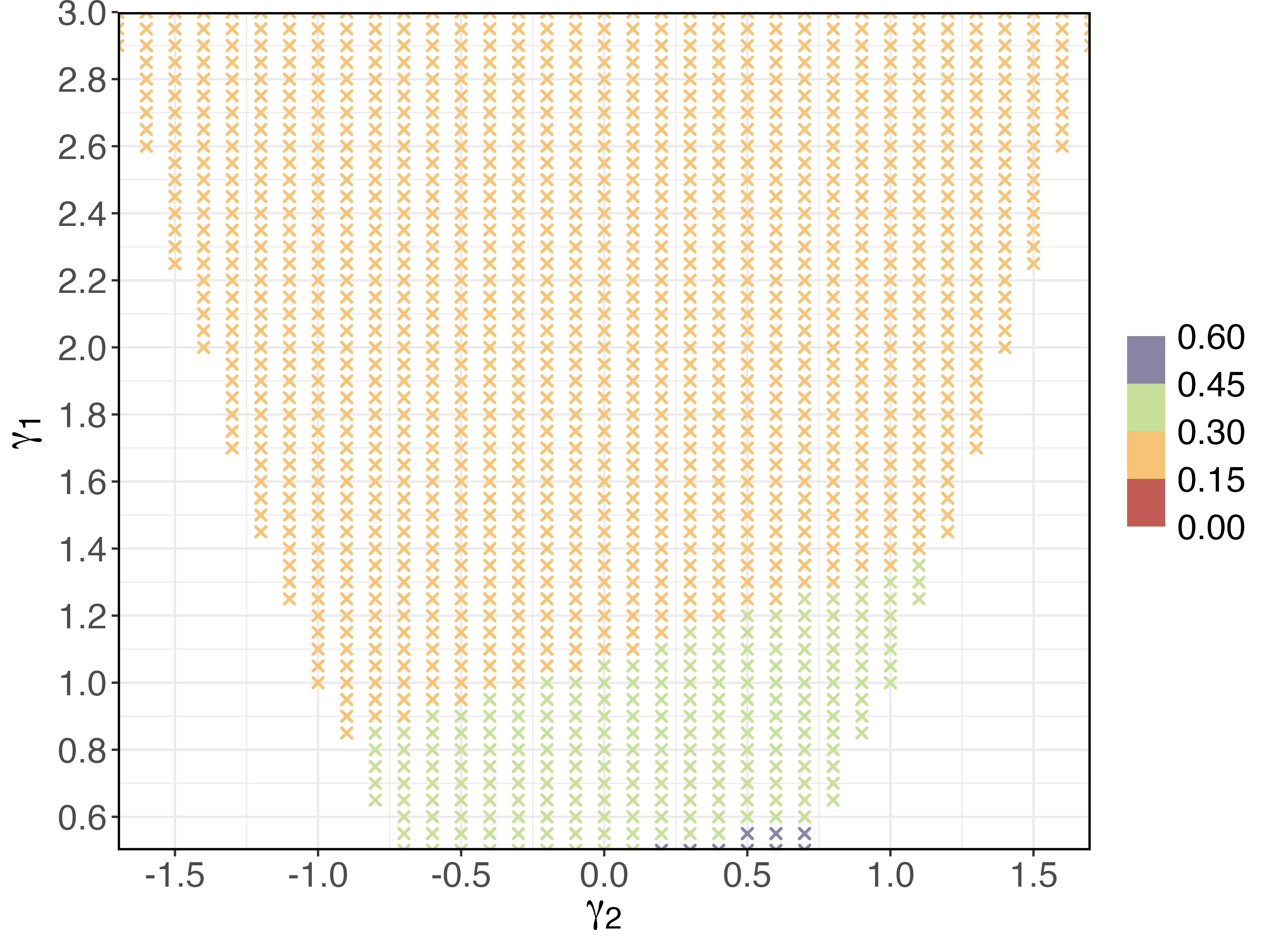}
        \caption{Point estimates of $\beta_{x\rightarrow y}$.}
        \label{fig:case_p_x2y_UV}
    \end{subfigure}
    \hfill
    \begin{subfigure}{0.49\linewidth}
        \centering
        \includegraphics[width=\linewidth]{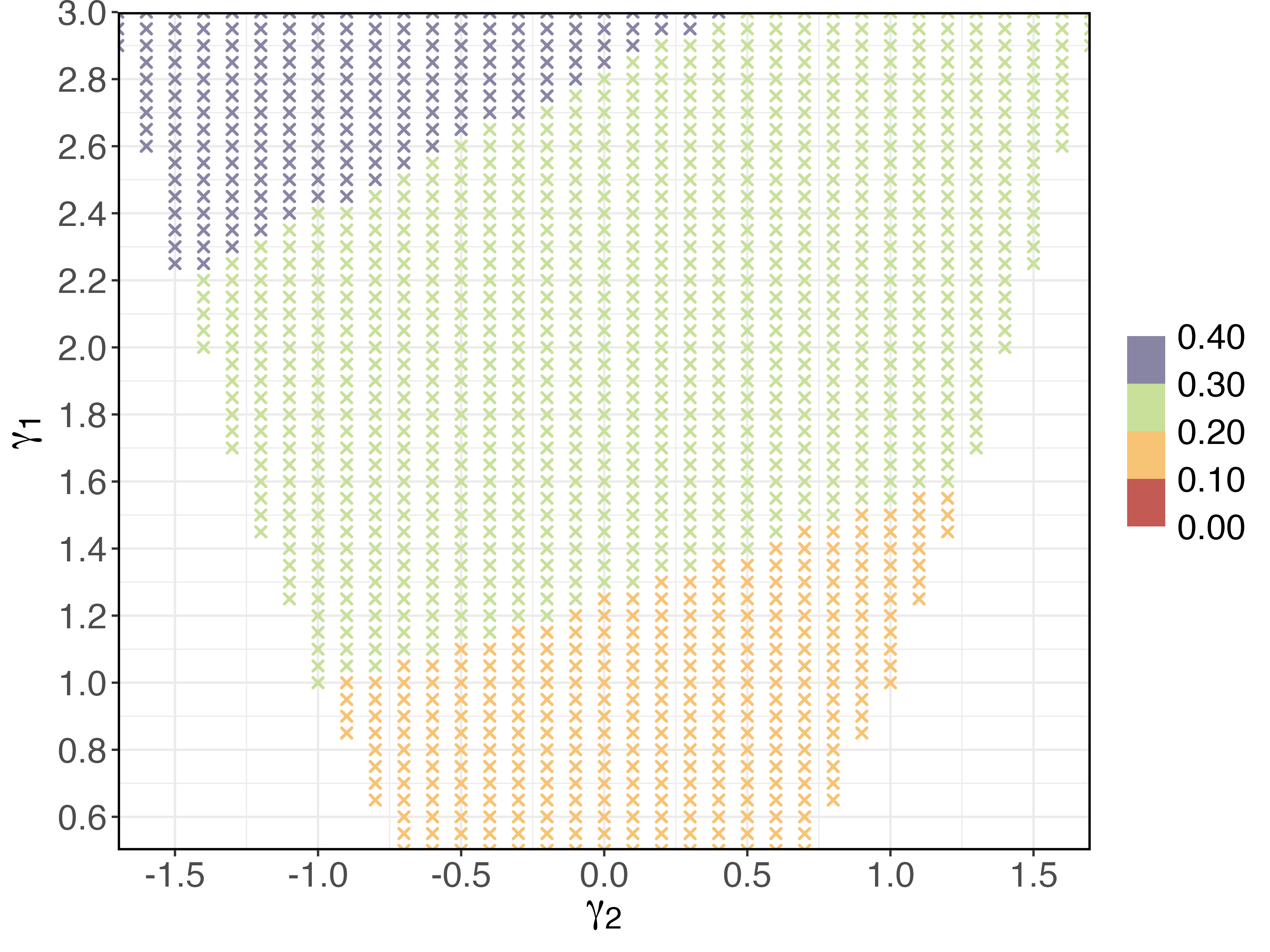}
        \caption{Point estimates of $\beta_{y\rightarrow x}$.}
        \label{fig:case_p_y2x_UV}
    \end{subfigure}

    \caption{Sensitivity analysis results for various parameter settings with the heart disease and diabetes dataset.}
    \label{fig:case_UV}
\end{figure}

In practice, it is challenging to determine the actual relative strength of unobserved confounding affecting the two target variables  -- heart disease'' and diabetes''. Therefore, parallel to Section \ref{ssec:UV_simulation}, we extend the study based on Assumption~\ref{assu:U} in Section~\ref{ssec:case estimate} by introducing parameters $\gamma_1$ and $\gamma_2$ to conduct a sensitivity analysis on this dataset. This aims to evaluate the accuracy of the proposed method in assessing possible real-world confounding scenarios. We vary $\gamma_1$ over $[0.5, 3]$ at intervals of 0.05, and $\gamma_2$ over $[-1.5, 1.5]$ at intervals of 0.1. { For each parameter combination $(\gamma_1,\gamma_2)$, we calculate the two candidate solution pairs given in Proposition~\ref{prop:UV} and retain those satisfying the corresponding sign restrictions. The resulting admissible estimates are displayed in Figure~\ref{fig:case_UV}.} Figure~\ref{fig:case_UV} uses color shading to represent the magnitudes and variations of these estimates.

As can be seen from Figure~\ref{fig:case_UV}, under different values of the parameter combinations $(\gamma_1,\gamma_2)$, the point estimates obtained from the empirical data are consistently positive. Moreover, according to the numerical range displayed in the figure, the range of the estimates when the absolute values of both parameters are close to zero is relatively consistent with the confidence interval obtained in Section~\ref{ssec:case estimate}. Overall, given the limited information on unobserved confounders, the method proposed in this paper can provide a reliable range for causal effect estimates, offering a practical reference value for real-world applications. Moreover, if more comprehensive information on confounders becomes available, such as knowledge from domain experts regarding the relative strength of the influence of unobserved confounders on heart disease and diabetes, the range of the causal effect point estimates can be further narrowed.

\subsection{Sensitivity analysis for ER assumption}\label{subsec:Sensitivity analysis for ER assumption}

In Sections~\ref{ssec:case estimate} and~\ref{ssec:case for UV}, we estimate the bidirectional causal effects between heart disease and diabetes and conduct corresponding sensitivity analyses under Assumption~\ref{assu:U} and its weaker version, respectively. These investigations are carried out within the framework of Model~\eqref{eq:bidirectional_model}, wherein the selected variables--``stroke" and ``BMI"--are assumed to be valid instrumental variables for heart disease and diabetes, respectively. Therefore, building upon Model~\eqref{eq:sensitivity-eq}, we further incorporate a sensitivity analysis addressing potential violations of the ER assumption for IVs in this dataset. This analysis follows the three corollaries presented in Section~\ref{ssec:sensitivity_iv}. Similar to the numerical simulation approach in Section~\ref{subsection:sensitivity analysis simulation}, the varying parameters across the three sensitivity parameter settings are all set to range from $[-0.16, 0.16]$ with an interval of 0.02.

\begin{figure}[h]
    \centering
\includegraphics[width=0.995\linewidth]{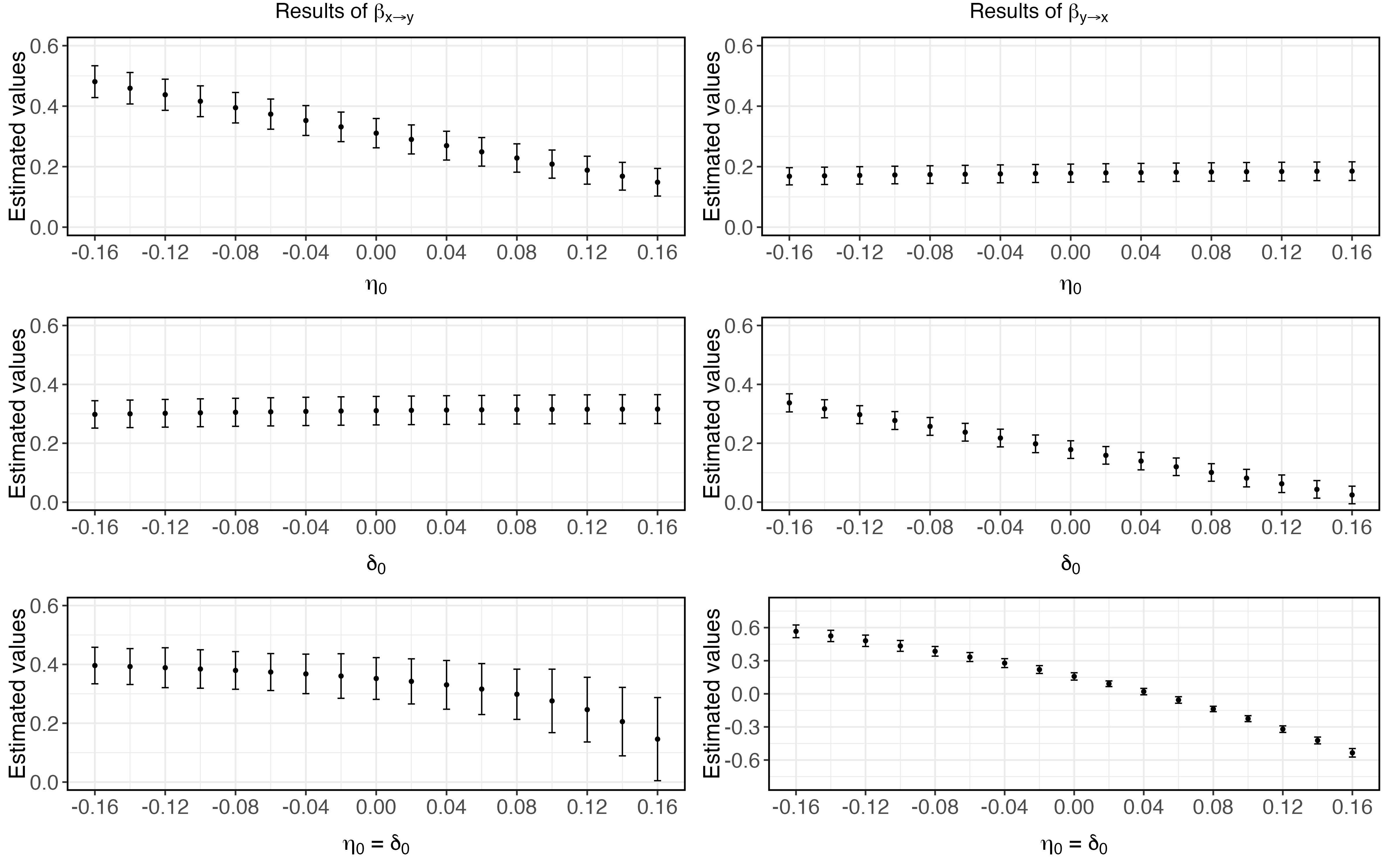}
    \caption{Sensitivity analysis results across three cases for the real dataset.}
    \label{fig:3case of iv for heart data}
\end{figure}

Figure~\ref{fig:3case of iv for heart data} presents the estimation results as error bar plots for three specific cases of sensitivity parameter values, with confidence intervals derived from 200 bootstrap samples. For Case 1, the sensitivity parameter $\eta_0$ has a significant impact on the causal relationship from $X$ to $Y$. As $\eta_0$ increases, the effect gradually weakens, but the causal effect remains positive throughout. Similarly, $\eta_0$ has a smaller impact on the reverse causal effect from $Y$ to $X$, and this effect also remains positive. For Case 2, the sensitivity parameter $\delta_0$ has a larger impact on the causal relationship from $Y$ to $X$. As $\delta_0$ increases, the effect gradually weakens, but remains positive. In contrast, $\delta_0$ has a smaller impact on the causal relationship from $X$ to $Y$, and the effect remains positive. For Case 3, the inclusion of two sensitivity parameters affects the causal relationships in both directions. The causal effect of $X$ on $Y$ gradually weakens as the sensitivity parameters increase, but remains positive throughout. In contrast, the causal effect of $Y$ on $X$ is more substantially influenced and turns negative when the sensitivity parameter exceeds 0.04. Introducing a sensitivity parameter $\eta_0$ (or $\delta_0$) can be interpreted as adding a causal path, which consequently influences the corresponding causal effect parameter $\beta_{x \to y}$ (or $\beta_{y \to x}$) to some extent.

Overall, whether under the change of $\eta_0$ or $\delta_0$, all bidirectional causal effects exhibit a positive relationship, suggesting a positive causal relationship between ``heart disease'' and ``diabetes'' based on real-world data, and the sensitivity analysis further validates the robustness of this relationship. Specifically, as the sensitivity parameters change, both the effect of heart disease on diabetes and its reverse both remain positive, indicating a significant bidirectional causal relationship. This relationship remains stable across a range of parameter values, providing reliable evidence supporting the causal relationship between heart disease and diabetes.

\section{Summary}\label{section:summary}

This paper develops a framework for evaluating bidirectional causal effects between binary variables under unobserved confounding by mapping continuous latent variables to observed binary outcomes via threshold constraints. We introduce instrumental variables to construct consistent estimators and propose a new identification strategy under the Probit model. We also relax the identification assumptions to varying degrees and conduct sensitivity analysis for the proposed method.

Several directions remain for future research. First, \citet{clarke2012instrumental} introduced a control function approach for binary outcomes, achieving identification of unidirectional causal effects under unobserved confounding. Extending this approach to bidirectional causal effects would be a promising direction. Second, since our method is designed for binary variables, a natural extension is to generalize the framework to multi-level settings using the ordered Probit model for categorical variables. These extensions are left for future work.


	\bibliographystyle{apalike}
	 
\bibliography{my}

\vskip .65cm
 
\vskip 1pt
\noindent
Shanshan Luo\\
 School of Mathematics and Statistics\\ Beijing Technology and Business University
\vskip 1pt
\noindent
E-mail: shanshanluo@btbu.edu.cn

\end{document}